\documentclass[12pt,showkeys,superscriptaddress,nofootinbib]{revtex4}
\usepackage{amsmath,amssymb,euscript,yfonts,psfrag,latexsym,dsfont,graphicx,bbm,color,amstext,wasysym,pdfsync,soul,framed}

\graphicspath{{/},{figures/}}

\newcommand{\E}{{\mathbb E}}
\newcommand{\D}{{\mathcal D}}
\newcommand{\A}{{\mathcal A}}
\newcommand{\B}{{\mathcal B}}

\newcommand{\bbP}{\mathbb{P}}
\newcommand{\mR}{\mathbb{R}}
\newcommand{\mN}{\mathbb{N}}

\newcommand{\mD}{{\mathbb D}}

\newcommand{\tr}{\operatorname{\rm trace}}
\newcommand{\diag}{\mathop{\mathrm{diag}}}

\definecolor{llgrey}{rgb}{0.9,0.9,0.9}
\definecolor{lgrey}{rgb}{0.6,0.6,0.6}
\definecolor{lred}{rgb}{0.9,0.7,0.7}

\newtheorem{thm}{Theorem}

\newtheorem{cor}{Corollary}

\newtheorem{prop}{Proposition}
\newtheorem{problem}{Problem}

\def \proof {\noindent {\em Proof. }}
\def \qed{\hfill $\Box$ \vskip 2ex}

\begin{document}

\title{Fast cooling for a system of stochastic oscillators}
\author{Yongxin Chen, Tryphon Georgiou and Michele Pavon
\thanks{Y. Chen and T. Georgiou are with the Department of Electrical and Computer Engineering,
University of Minnesota,
200 Union street S.E. Minneapolis, Minnesota MN 55455, U.S.A. {\tt\small chen2468@umn.edu}, {\tt\small tryphon@umn.edu}}
\thanks{Michele Pavon is with the Dipartimento di Matematica,
Universit\`a di Padova, via Trieste 63, 35121 Padova, Italy {\tt\small pavon@math.unipd.it}}}

\begin{abstract}
We study feedback control  of coupled nonlinear stochastic oscillators in a force field. We first consider the problem of asymptotically driving the system to a desired {\em steady state} corresponding to reduced thermal noise.
Among the feedback controls achieving the desired asymptotic transfer, we find that the most efficient one {from an energy point of view} is characterized by  {\em time-reversibility}. We also
extend the theory of Schr\"{o}dinger bridges to this model, thereby steering the system in {\em finite time} and with minimum effort to a target steady-state distribution. The system can then be maintained in this state
through the optimal steady-state feedback control. The solution, in the finite-horizon case,  involves a space-time harmonic function $\varphi$, and $-\log\varphi$ plays the role of an artificial, time-varying potential in which the desired evolution occurs. This framework appears extremely general and flexible and can be viewed as a considerable generalization of existing active control strategies such as macromolecular cooling. In the case of a quadratic potential, the results assume a form particularly attractive from the algorithmic viewpoint as the optimal control can be computed via deterministic matricial differential equations. An example involving inertial particles illustrates both transient and steady state optimal feedback control.
\end{abstract}

\keywords
{Stochastic oscillator, steady-state, cooling, Schr\"{o}dinger bridges, stochastic control, reversibility.}
\maketitle
\section{Introduction}\label{sec:introduction}
Stochastic oscillators represent a most fundamental model of dissipative processes since the 1908 paper by Paul Langevin \cite{Lan} which appeared three years after the ground-breaking work of Einstein and Smoluchowski. These stochastic models culminated in 1928 in the Nyquist-Johnson model for RLC networks with noisy resistors and in 1930 in the Ornstein-Uhlenbeck model of physical Brownian motion \cite{N1}. In more recent times, they  play a central role in {\em cold damping feedback}. The latter is employed to reduce the effect of thermal noise on the motion of an oscillator by applying a viscous-like force, which is historically one of the very first feedback control actions ever analyzed\footnote{``In one class of regulators of machinery, which we may call {\em moderators}, the resistance is increased by a quantity depending on the velocity", James Clerk Maxwell, On Governors, {\em Proceedings of the Royal Society}, no. {\bf 100} (1868),  270-282.}. It was first implemented in the fifties on electrometers \cite{MZV}. Since then, it has been successfully employed in a variety of areas such as  atomic force microscopy (AFM) \cite{LMC},  polymer dynamics \cite{DE,BBP} and nano to
meter-sized resonators, see \cite{FK,MG,SR,Vin,PV}. These new applications also pose new {\em physics} questions as the system is driven to a non-equilibrium {\em steady state} \cite {Q,KQ,BCD,PT1}. In \cite{FH}, a suitable {\em efficiency measure} for these diffusion-mediated devices was introduced which involves a class of stochastic control problems.

In spite of the flourishing of these applications and cutting edge developments, the interest in these problems in the control engineering community has been shallow to say the least.
However, 
as we argue below, 
these problems may be cast in the framework of a suitable extension of the classical theory of Schr\"{o}dinger bridges for diffusion processes \cite{W} where the time-interval is {finite or} infinite.
Moreover, a connection between {finite-horizon Schr\"{o}dinger bridges} and the so called ``logarithmic transformation" of stochastic control of Fleming, Holland, Mitter  {\em et al.}, see e.g.\cite{F}, has been known for some time \cite {DP,Bl,DPP,PW}.
Excepting some special cases \cite{FH,FHS}, however, the optimal control is not provided by the theory in an {\em implementable form} and a wide gap persists between the simple constant linear feedback controls used in the laboratory and the Schr\"{o}dinger bridge theory which requires the solution of two partial differential equations nonlinearly coupled through their boundary values --these coupled differential equations are known as a ``Schr\"odinger system'' \cite{W}. Only 
recently some
progress has been made  in deriving implementable forms of the optimal control for general linear stochastic systems  \cite{CG2014,CGP,CGP_ACC15,CGP2}
as well as implementable solutions of analogous Schr\"{o}dinger systems for Markov chains, Kraus maps of statistical quantum mechanics, and for diffusion processes \cite{GP,CGP6.15}.

In this paper, continuing the work of \cite{CG2014,CGP,CGP_ACC15,CGP2}, we study a general system of nonlinear stochastic oscillators. For this general model, we prove optimality of certain feedback controls which are given in an explicit or computable form. We also highlight the relevance of optimal controls on examples of stochastic oscillators. 
{In Section \ref{sec:system} we introduce the system of nonlinear stochastic oscillators and discuss a fluctuation-dissipation relation and reversibility. In Section \ref{invariant} we discuss thoroughly the existence of invariant measures and related topics such as ergodicity and convergence to equilibrium first in the case of linear dynamics and then in the general case. In Section \ref{sec:optimal}, we characterize the most efficient feedback law which achieves the desired asymptotic cooling and relate optimality to reversibility of the controlled evolution. In Section \ref{sec:cooling}, we show how the desired cooling can be accomplished in finite time using a suitable generalization of the theory of Schr\"{o}dinger bridges. The latter results are then specialized in the following section, Section \ref{sec:quadratic}, to the case of a quadratic potential where the equations become linear and the results of \cite{CGP} lead to implementable optimal controls. Optimal transient and steady state feedback controls are illustrated in one example involving inertial particles in Section \ref{sec:example}.}

\section{A system of stochastic oscillators}\label{sec:system}
Consider a mechanical system in a force field
coupled to a heat bath. More specifically,  consider the following generalization of the Ornstein-Uhlenbeck
model of physical Brownian motion \cite{N1}
\begin{subequations}\label{OU}
\begin{eqnarray}\label{OU1}
dx(t)&=&v(t)\,dt,\quad x(t_0)=x_0 \;{\rm a.s.}\\\label{OU2}
Mdv(t)&=&-Bv(t)\,dt-\nabla_x V(x(t))dt+\Sigma dW(t), \quad v(t_0)=v_0 \; {\rm a.s.}
\end{eqnarray}
\end{subequations}
that was also studied 
in \cite{HP1}.
Here $x(t)$ and $v(t)$ take values in $\mR^n$ where $n=3N$ and $N$ is the number of oscillators. The potential $V\in C^1$ (i.e., continuously differentiable), is bounded below and tends to infinity
for $\|x\|\rightarrow\infty$. 
The noise process $W(\cdot)$ is a standard $n$-dimensional Wiener process independent of the pair $(x_0,v_0)$. The matrices $M$, $B$ and $\Sigma$ are $n\times n$ with $M$  symmetric and positive definite, and $\Sigma$ nonsingular. We also assume throughout the paper that  $B+B'$, where $'$ denotes the transposition, is positive semi-definite. 

The one-time phase space probability density $\rho_t(x,v)$, or more generally probability measure\footnote{When  $\mu_t(x,v)$ is absolutely continuous, $\mu_t(x,v)(dxdv)=\rho_t(x,v)dxdv$.} $\mu_t(x,v)$, represents the state of the thermodynamical system at time $t$. Notice that we allow for both potential and dissipative interaction among the particles/modes, with velocity coupling and with dissipation described by a linear law. The models that will be discussed in Section \ref{sec:cooling} are more special and correspond to the situation where $M$, $B$ and $\Sigma$ are in fact diagonal matrices.
Other spatial
arrangements and interaction patterns may be
accommodated in this frame as, for instance, 
a ring of $N$-oscillators  with $x_0=x_N$ described by the scalar equations
\begin{subequations}
\label{eq:5}
\begin{eqnarray}\label{ring1} dx_k&=&v_kdt,\\\label{ring2} m_kdv_k&=&\left(-\gamma v_{k-1}-\beta v_k-\gamma v_{k+1}-\frac{\partial V(x)}{\partial x_k}\right)dt + \sigma_k dW,
\end{eqnarray}
\end{subequations}
where $\sigma_k\in\mR^{1\times n}$, cf.\ \cite[Section 6]{HP1}. 
For this case, 
(\ref{eq:5}) can be put in the form (\ref{OU}) by defining
$$M=\diag (m_1,\ldots,m_N),\quad
B=\left(\begin{matrix}\beta& \gamma& 0&0&\cdot & \gamma\\
\gamma& \beta& \gamma& 0&\cdot &0\\
0&\gamma&\beta&\gamma&\cdot &0\\
0&0&\gamma&\beta&\cdot&\cdot\\
\cdot&\cdot&\cdot&\cdot&\cdot&\gamma\\
\gamma&\cdot&\cdot&\cdot&\gamma&\beta\end{matrix}\right),
\quad \Sigma=\left(\begin{matrix}\sigma_1\\\cdot\\\cdot\\\cdot\\\sigma_N\end{matrix}\right).
$$
Besides thermodynamics, applications of this basic model of dissipative processes
is found in nonlinear
circuits with noisy resistors, in chemical physics, in biology, and other fields, e.g., see  \cite{TW,LE,RS}.

\subsection{Boltzmann's distribution and a fluctuation-dissipation relation}
According to the Gibbsian postulate of classical statistical
mechanics, the {\em equilibrium state} of a
system in contact with a heat bath at constant absolute temperature $T$ and with
Hamiltonian function $H$ is necessarily given by the
Boltzmann distribution law
\begin{equation}\label{B}\rho_{B}=Z^{-1}\exp\left[-\frac{H}{kT}\right]
\end{equation}
where $Z$ is the {\em partition function}\footnote{We assume here and throughout the paper that $V$ is such that $\exp\left[-\frac{H}{kT}\right]$ is integrable on $\mR^n\times\mR^n$.}. The Hamiltonian function corresponding to
 (\ref{OU}) is
$$H(x,v)=\frac{1}{2}\langle v,Mv\rangle+V(x),
$$
where $\langle \cdot,\cdot\rangle$ denotes the Euclidean scalar
product in $\mR^n$; the partition function is simply a
normalization constant.

The key mathematical concept relevant to a stochastic characterization of equilibrium is that of an {\em invariant probability measure}. 
However, not all invariant probability measures correspond to equilibrium. They may represent a {\em steady state} of nonequilibrium thermodynamics. Thus, while it is important to establish existence and uniqueness of the invariant probability measure, it is also necessary to characterize when we can expect such a measure to be of the Boltzmann-Gibbs type (\ref{B}). For the system of stochastic oscillators (\ref{OU}), this was established in \cite{HP1}, generalizing the
Einstein {\em fluctuation-dissipation relation}:
\begin{prop}\label{Pro1} An invariant  measure for {\em (\ref{OU})} is a Boltzmann distribution with density
{\em (\ref{B})}   if
and only if
\begin{equation}\label{FD}
\Sigma\Sigma^\prime=kT(B+B').
\end{equation}
\end{prop}
Before dealing with existence of invariant measures, we discuss reversibility.

\subsection{Reversibility}\label{reversibility}

Let us start recalling that a stochastic process $\{X(t), t\in T\}$ taking values in $\cal X$ and with the invariant measure $\bar{\mu}$ is called reversible if its finite dimensional distributions coincide with those of the time-reversed process. Namely, for all $t_1<t_2<\cdots<t_m$ and $x_i\in{\cal X}$,
$$\bbP_{\bar{\mu}}(X(t_1)=x_1,X(t_2)=x_2,\ldots,X(t_n)=x_n)=\bbP_{\bar{\mu}}(X(t_1)=x_n,X(t_2)=x_{n-1},\ldots,X(t_n)=x_1).
$$

For a Markov-diffusion process such as (\ref{OU}), it should be possible to characterize reversibility through the stochastic differentials. Indeed, it has been shown by Nelson \cite{N1,N2}, see also \cite{HP}, that Markov diffusion processes admit, under rather mild conditions, a reverse-time stochastic differential. For (\ref{OU}), this stochastic differential takes the form
\begin{subequations}
\begin{eqnarray}\label{Rev1}
dx(t)&=&v(t)\,dt,\\\label{Rev2}
Mdv(t)&=&-Bv(t)\,dt-\nabla_x V(x(t))dt-\Sigma\Sigma'M^{-1}\nabla_v\log\rho_t(x(t),v(t))+\Sigma dW_-(t).
\end{eqnarray}
\end{subequations}
Here $dt>0$, $\rho_t$ is the probability density of the process in phase space, and $W_-$ is a standard Wiener process whose past $\{W_-(s);0\le s\le t\}$ is independent of $ \left(\begin{matrix}x(t)\\v(t)\end{matrix}\right)$ for all $t\ge 0$.

Consider now the situation where $\rho_t(x,v)=\bar{\rho}(x,v)$ an invariant density. Consider also the time reversal transformation \cite{Gu}
$$
t\rightarrow t'=-t,\quad x\rightarrow x'=x,\quad v\rightarrow v'=-v,\quad \nabla_x\rightarrow\nabla_{x'}=\nabla_x,\quad \nabla_v\rightarrow\nabla_{v'}=-\nabla_v.
$$
In view of (\ref{OU2}) and (\ref{Rev2}), we also define
$$F_+(x,v,t)=-Bv-\nabla_xV(x),\quad F_-(x,v,t)=-Bv-\nabla_x V(x)-\Sigma\Sigma'M^{-1}\nabla_v\log\bar{\rho}(x,v).
$$
Then, we have invariance under time reversal if and only if
\begin{eqnarray}\nonumber F_+'(x',v',t')=F_-(x,v,t)=-Bv-\nabla_x V(x)-\Sigma\Sigma'M^{-1}\nabla_v\log\bar{\rho}(x,v)\\=F_+(x',v',t')=-Bv'-\nabla_{x'} V(x')=Bv-\nabla_x V(x).\nonumber
\end{eqnarray}
We get the condition
\begin{equation}\label{TRI}
\Sigma\Sigma'M^{-1}\nabla_v\log\bar{\rho}(x,v)=-2Bv.
\end{equation}
We have therefore the following result.
\begin{prop}\label{Pro2} The phase-space  process (\ref{OU}) with the invariant Boltzmann distribution (\ref{B}) is reversible if and only if the matrix $B$ is symmetric positive definite. 
\end{prop}
\proof Since $\nabla_v\log\rho_B(x,v)=-\frac{1}{kT}Mv$, (\ref{TRI}) reads
$$\frac{1}{kT}\Sigma\Sigma'v=2Bv, \quad v\in\mR^n.
$$
which holds true if and only if $B$ is symmetric positive definite ($\Sigma$ is nonsingular) satisfying (\ref{FD}), namely
\begin{equation}\label{FD'}
\Sigma\Sigma'=2kTB.
\end{equation}

\qed
In \cite[Proposition 2.1]{HP1}, it was shown that, under (\ref{FD}), symmetry of $B$ is a necessary and sufficient condition for a Newton-type law to hold. The latter can be derived from a Hamilton-like principle in analogy to classical mechanics \cite{P}.
In the next section, we deal in some detail with the issue of existence and properties of an invariant measure for (\ref{OU}).

\section{Invariant measures for the system of stochastic oscillators}\label{invariant}
This topic is in general a rather delicate one and the mathematical literature covering model (\ref{OU}) is rather scarce. We have therefore decided to give a reasonably comprehensive account of the issues and results. We discuss first the case of a quadratic potential where the dynamics becomes linear and simple linear algebra conditions may be obtained. This case is also of central importance for  cooling applications \cite{LMC,DE,Vin}.
\subsection{Invariant measures: The case of a quadratic potential function}\label{quadratic}
We assume in this subsection that
$$V(x)=\frac{1}{2}\langle x,Kx\rangle
$$
with $K$ symmetric positive definite so that the various restoring forces in the vector Langevin equation (\ref{OU}) are {\em linear} and the system takes the form: 
\begin{subequations}
\begin{equation}\label{OUL}
d\xi
={\cal A}\xi
dt+{\cal B}dW(t)
\end{equation}
where
\begin{equation}\label{AB}\xi=\left(\begin{matrix}x\\v\end{matrix}\right),\quad \cal A=\left(\begin{matrix}0&I\\-M^{-1}K&-M^{-1}B\end{matrix}\right), \quad \cal B=\left(\begin{matrix}0\\M^{-1}\Sigma\end{matrix}\right).
\end{equation}
\end{subequations}
This case has been thoroughly studied in \cite[Section 5]{HP1} building on the deterministic results of M\"{u}ller \cite{Mu} and Wimmer \cite{Wim}. Thus, we only give below the essential concepts and results for the sake of continuity in exposition.

As is well known \cite{FCG}, the existence of a {\em Gaussian} invariant measure with nonsingular covariance matrix ${\cal P}$ is intimately connected to the existence of a positive definite ${\cal P}$ satisfying the {\em Lyapunov equation}
\begin{equation}\label{Lyap}
0=\cal A\cal P+\cal P\cal A'+\cal B\cal B'.
\end{equation}
Inertia theorems for (\ref{Lyap}) \cite{Wim} relate the spectrum of ${\cal P}$ to the spectrum of ${\cal A}$ and {\em controllability} of an associated deterministic system. Recall that for a dynamical system 
$$\dot{\xi}(t)=f(\xi(t),u(t),t), \quad \xi(0)=\xi_0$$ 
(complete) controllability refers to the property of an external input (the vector of control variables $u(t)$) to steer the internal state $\xi(t)$ in finite time from any initial condition $\xi_0$ to any desired target state. It turns out that the pair $({\cal A},{\cal B})$ gives rise to a controllable linear  system 
$$\dot{\xi}(t)={\cal A}\xi(t)+{\cal B}u(t)$$ 
if and only if the matrix $({\cal B},{\cal A}{\cal B},\ldots, {\cal A}^{2n-1}{\cal B})$ has full row rank \cite{KFA}. Now, suppose $\cal P$ is positive definite and satisfies (\ref{Lyap}). Let $\lambda$ be an eigenvalue of $\cal A'$ with $\gamma$ a corresponding eigenvector. Then
$$0=\gamma'\left[\cal A\cal P+\cal P\cal A'+\cal B\cal B'\right]\gamma=(\lambda+\bar{\lambda})\gamma'\cal P\gamma+\gamma'\cal B\cal B'\gamma.
$$
Since $\gamma'{\cal P}\gamma>0$ and $\gamma'{\cal B}{\cal B}'\gamma\ge 0$, it follows that $\Re[\lambda]\le 0$.
That is, the spectrum of ${\cal A}$ is contained in the left half of the complex plane. In the other direction,
if ${\cal A}$ is asymptotically stable (i.e., all eigenvalues are in the {\em open} left half-plane), $\cal P$ given by
$${\cal P}=\int_0^\infty e^{{\cal A}\tau}{\cal B}{\cal B}'e^{{\cal A}'\tau}d\tau
$$
satisfies (\ref{Lyap}) and is positive semidefinite --this is the so-called {\em controllability Gramian}. It turns out that this is positive definite if and only if the pair $({\cal A},{\cal B})$ is controllable \cite{FCG}.

For $({\cal A},{\cal B})$ as in (\ref{AB}) and under the present assumptions ($\Sigma$ nonsingular), 
the matrix $({\cal B},{\cal A}{\cal B},\ldots, {\cal A}^{2n-1}{\cal B})$ always has full row rank. Thus, existence and uniqueness of a nondegenerate Gaussian invariant measure is reduced to characterizing asymptotic stability of the matrix ${\cal A}$ in (\ref{AB}).  When ${\cal A}$ is asymptotically stable, starting from any initial Gaussian distribution, we have convergence to the invariant Gaussian density with zero mean and covariance ${\cal P}$. Asymptotic stability of ${\cal A}$ can be studied via stability theory for the deterministic system
$$M\ddot{z}(t)+\frac{B+B'}{2}\dot{z}(t)+Kz(t)=0
$$
employing as Lyapunov function the energy $H(x,v)=\frac{1}{2}\langle v,Mv\rangle+\frac{1}{2}\langle x,Kx\rangle$. In the case when $B+B'$ is positive semidefinite, using invariance of controllability under feedback, the asymptotic stability of ${\cal A}$ was shown by M\"{u}ller \cite{Mu} to be equivalent to the complete
controllability of the system
\begin{equation}\label{deterministic}
M\ddot{z}(t)+Bu(t)+Kz(t)=0.
\end{equation}
In M\"{u}ller's terminology, as quoted in \cite{Wim}, this means that
damping in  the corresponding deterministic system is {\em pervasive}. We collect all these findings in the following theorem.
\begin{thm} {\em \cite{Mu,Wim}} In model {\em (\ref{OU})}, assume that $M=M'>0$, $V(x)=\frac{1}{2}\langle x,Kx\rangle$ with $K=K'>0$. Suppose moreover that $(B+B')\ge 0$ and that $\Sigma$ is nonsingular. Then there exists a unique nondegenerate invariant Gaussian measure if and only if the pair of matrices
\begin{equation}
\left(\left[\begin{matrix}0&I\\-M^{-1}K&0\end{matrix}\right], \left[\begin{matrix}0&0\\0&-M^{-1}\frac{B+B'}{2}\end{matrix}\right]\right)
\end{equation}
is controllable. In particular, this is always the case when $B+B'$ is actually positive definite. If the invariant measure exists, it is of the Boltzmann type {\em (\ref{B})} if and only if the generalized fluctuation-dissipation relation {\em (\ref{FD})} holds.
\end{thm}
Some extensions of this result to the case of a non quadratic potential have been presented in \cite[Section 3B]{BM}.

\subsection{Invariant measures: The case of a general potential function}\label{generalcase}

Consider now the general case where the potential function $V$ is any nonnegative, continuously differentiable function which tends to infinity for $\|x\|\rightarrow\infty$. As already observed, existence, uniqueness, ergodicity, etc.\ of an invariant probability measure are quite delicate issues and we refer to the specialized literature for the full story, see e.g. \cite[Section 7.4]{St}, \cite[Chapters 5 and 7]{DaPrato}. One way to prove existence of an invariant measure is by establishing that the flow of one-time marginals $\mu_t(x_0,v_0)$, $t\ge 0$ of the random evolution in (\ref{OU}) starting from the point $(x_0,v_0)$ is {\em tight}\footnote{A set $\Lambda$ of probability measures $\mu$ on $\mR^m$ is called tight if for every $\epsilon>0$ there exists a compact set $C_\epsilon\subset\mR^m$ such that for any $\mu\in\Lambda$ it holds $\mu(C_\epsilon)\ge 1-\epsilon$. If the family $\mu_t(x_0,v_0)$ is tight, one can, by Prokhorov's theorem \cite[Theorem 6.7]{DaPrato},  extract a weakly convergent sequence $\mu_{t_n}(x_0,v_0), n\in\mN$. A sequence $\mu_n$ converges weakly to $\mu$ (one writes $\mu_n\rightharpoonup \mu$) if $\int\varphi\mu_n\rightarrow\int\varphi\mu$ for every bounded, continuous function $\varphi$.}. If that is the case, existence of an invariant measure follows from the Krylov-Bogoliubov theorem \cite[Section 7.1]{DaPrato}. One way to establish tightness of the family $\mu_t(x_0,v_0), t\ge 0$ is via Lyapunov functions.  One has, for instance, the following result. 
\begin{prop}\label{existence}{\em  \cite[Proposition 7.10]{DaPrato}} Let ${\cal V}:\mR^{2n}\rightarrow [0,+\infty]$ be a Borel function whose level sets $K_a=\{(x,v)\in\mR^{2n}: {\cal V}(x,v)\le a\}$ are compact for all $a>0$. Suppose there exists $(x_0,v_0)\in\mR^{2n}$ and $C(x_0,v_0)>0$ such that the corresponding solution $(x(t,x_0),v(t,v_0))$ of (\ref{OU}) starting from $(x_0, v_0)$ is such that
\begin{equation}
\E\{{\cal V}(x(t,x_0),v(t,v_0))\}\le C(x_0,v_0),\quad \forall t\ge 0.
\end{equation}
Then, there exists an invariant measure for (\ref{OU}).
\end{prop}

The natural Lyapunov function for our model is the Hamiltonian $H(x,v)$ which, under the present assumptions on the potential function $V$, does have compact level sets.
Thus, we
now consider the evolution of $H(x(t),v(t))$ along the random evolution of (\ref{OU}). By Ito's rule \cite{KS}, we get
\begin{eqnarray}\nonumber
dH(x(t),v(t))&=&\nonumber\left[\frac{\partial H}{\partial t}+\left(\begin{matrix}v(t)\\\left(-M^{-1}Bv(t)-M^{-1}\nabla_x V(x(t)\right)\end{matrix}\right)\cdot\left(\begin{matrix}\nabla_x H\\\nabla_v H\end{matrix}\right)\right.\\\nonumber
&&\left.+\frac{1}{2}\sum_{i,j=1}^n\left[M^{-1}\Sigma\Sigma'M^{-1}\right]_{ij}\frac{\partial^2 H}{\partial v_i\partial v_j}\right](x(t),v(t))dt\\\nonumber&&+\nabla_v H(x(t),v(t))\cdot M^{-1}\Sigma dW(t)\\&=&-\langle Bv(t),v(t)\rangle dt+\frac{1}{2}\tr\left[M^{-1}\Sigma\Sigma'\right]dt+v(t)'\Sigma dW(t).\label{diffenergy}
\end{eqnarray}
Let $U(t)=\E\{ H(x(t),v(t))\}$ be the {\em internal energy}. Then from (\ref{diffenergy}), observing that $\langle Bv,v\rangle=\langle B'v,v\rangle$, we get
\begin{equation}\label{first law}
U(t+h)-U(t)=\E\left\{\int_t^{t+h}-\left\langle \frac{B+B'}{2}v(\tau),v(\tau)\right\rangle d\tau \right\}+\frac{h}{2}\tr\left[M^{-1}\Sigma\Sigma'\right].
\end{equation}
The first term represents the {\em work} done on the system by the friction forces, whereas the second is due to the action of the thermostat on the system and represents the {\em heat}, so that (\ref{first law}) appears as an instance of the first law of thermodynamics
\[\Delta U=W+Q.
\] 
Since the friction force is dissipative ($B+B'\ge 0$), we have that $W\le 0$.  If we take  $(0,0)$ as initial condition for (\ref{OU}), the initial variance will be zero and therefore by (\ref{first law}) the internal energy will initially increase. The statement 
that it remains bounded, so that we can apply Proposition \ref{existence}, rests on the possibility that the $L^2$ norm of $v(\tau)$, suitably weighted by the symmetric part of the friction matrix $B$, becomes eventually at least as large as the constant quantity $\tr\left[M^{-1}\Sigma\Sigma'\right]$. 

In the rest of this section, we discuss the case where the generalized fluctuation-dissipation relation (\ref{FD}) holds. Then, a direct computation on the Fokker-Planck equation associated to (\ref{OU})
\begin{equation}\label{FP}
\frac{\partial{\rho}}{\partial{t}} +v\cdot \nabla_x\rho
-\nabla_v\cdot\left(M^{-1}Bv+M^{-1}\nabla_x V\rho\right) =
\frac{1}{2}\sum_{i,j=1}^n\left[M^{-1}\Sigma\Sigma'M^{-1}\right]_{ij}\frac{\partial^2}{\partial v_i\partial
v_j}\rho
\end{equation}
shows that the Boltzmann density (\ref{B})
$$\rho_{B}(x,v)=Z^{-1}\exp\left[-\frac{H(x,v)}{kT}\right]=Z^{-1}\exp\left[-\frac{\frac{1}{2}\langle v,Mv\rangle+V(x)}{kT}\right]$$ 
is indeed invariant. We now discuss uniqueness, ergodicity and convergence of $\rho_t$ to $\rho_B$.
Consider the  {\em free energy} functional
$$F(\rho_t)=kT \int_{\mR^{n}}\int_{\mR^{n}}\log\frac{\rho_t}{\rho_B}\rho_t
\,dxdv=kT \,\mD(\rho_t\|\rho_B),
$$
where $\mD(\rho\|\sigma)$ is the {\em relative entropy} or {\em divergence} or {\em Kullback-Leibler pseudo-distance} between the densities $\rho$ and $\sigma$. We have the well known
result, see e.g. \cite{Gr}:

\begin{equation}\label{FE}\frac{d}{dt}F(\rho_t)=-\frac{kT}{2}
\int_{\mR^{n}}\int_{\mR^n}\langle
\Sigma\Sigma^\prime \nabla_v\log\frac{\rho_t}{\rho_B},\nabla_v\log\frac{\rho_t}{\rho_B}\rangle\rho_t\,dxdv
\end{equation}
Recalling that $\mD(\rho\|\sigma)\ge 0$ and $\mD(\rho\|\sigma)=0$ if and only if $\rho=\sigma$ \cite{Ku}, we see that $F$ acts as a natural Lyapunov function for (\ref{FP}). The decay of $F(\rho_t)$ implies uniqueness of the invariant density $\rho_B$ in the set $\{\rho | \mD(\rho\|\rho_B)<\infty\}$. Suppose now that $V$ is actually $C^\infty$. Then the {\em generator} (\cite{KS})
 of (\ref{OU}), taking $M=\Sigma\Sigma'=I$ to simplify the writing,
\begin{equation}\label{generator} 
v\cdot\nabla_x-Bv\cdot\nabla_v-\nabla_x V\cdot\nabla_v+\frac{1}{2}\Delta_v
\end{equation}
is actually {\em hypoelliptic} \cite{Bell}. Indeed it can be written in H\"{o}rmander's form
$$X_0+Y+\frac{1}{2}\sum_{i=1}^nX_i^2
$$
where 
$$X_0=-Bv\cdot\nabla_v, \quad Y=-\nabla_x\cdot\nabla_v+v\cdot\nabla_x, \quad X_i=\frac{\partial}{\partial v_i}.$$
Moreover, the vectors
$$(X_1,\ldots,X_n,[Y,X_1],\ldots,[Y,X_n])
$$
form a basis of $\mR^{2n}$ at every point  \cite[Section 2]{Bau}. This is H\"{o}rmander's condition \cite{Ho} which, in the case of a quadratic potential, turns into controllability of the pair $({\cal A},{\cal B})$ in (\ref{OUL}). It follows that, for any initial condition $\rho_0$ (even a Dirac delta) the correspondig solution $\rho_t$ of (\ref{FP}) is smooth and supported on all of $\mR^{2n}$ for all $t>0$.  Let $p(s,\xi,t,\eta)$ denote the (smooth) transition density and consider the Markov semigroup
$$P_t[\varphi](\xi)=\int p(0,\xi,t,\eta)\varphi(\eta)d\eta,
$$
for $\varphi$ a Borel bounded function on $\mR^{2n}$. Then the Markov semigroup is {\em regular}  \cite[Definition 7.3]{DaPrato} and the invariant measure $\rho_B(x,v)dxdv$  is unique \cite[Proposition 7.4]{DaPrato}. This invariant measure being unique, it is necessarily ergodic \cite[Theorem 5.16]{DaPrato} (time averages converging to probabilistic averages).

We finally turn to the convergence of $\rho_t$ to $\rho_B$.  In view of (\ref{FE}), it seems reasonable to expect that  $\rho_t(x,v)$ tends to $\rho_B(x,v)$ in {\em relative entropy} and, consequently, in $L^1$ (total  variation of the measures) \cite{LC}. This, however, does not follow from (\ref{FE}) and turns out to be surprisingly difficult to prove. Indeed, the result rests on the possibility of establishing a {\em logarithmic Sobolev inequality} (LSI) \cite{Le,Bau}, \cite[Section 9.2]{Vil}, a topic which has kept busy some of the finest analysts during the past forty years.  One says the probability measure $\mu$ satisfies a (LSI) with constant $\lambda>0$ if for every function $f$ satisfying $\int f^2d\mu=1$, 
\begin{equation}\label{LSI}
\int f^2\log f^2 d\mu\le \frac{1}{2\lambda}\int\|\nabla f\|^2d\mu.
\end{equation}
Let us consider a non degenerate diffusion process $\{X(t); t\ge 0\}$ taking values in some Euclidean space $\mR^m$ with differential
$$dX(t)=-\frac{1}{2}\nabla R(X(t))dt+dW(t),
$$
where $R$  is a smooth, nonnegative function such that $\exp[-R(x)]$ is integrable over $\mR^m$. Then $\rho_\infty(x)=C\exp[-R(x)]$ is an invariant density for $X(t)$ where $C$ is a normalizing constant. Let $\rho_t$ be the one-time density of $X(t)$. Then, in analogy to (\ref{FE}), we have the decay of the relative entropy
\begin{equation}\label{Decay}
\frac{d}{dt}\mD(\rho_t\|\rho_\infty)=-\frac{1}{2}\int\|\nabla\log\frac{\rho_t}{\rho_\infty}\|^2\rho_t dx.
\end{equation}
The integral appearing in the right hand-side of (\ref{Decay}) is called the {\em relative Fisher information} of $\rho_t$ with respect to $\rho_\infty$. It is also 
a ``Dirichlet form", as it can be rewritten as
$$4\int \|\nabla \sqrt{\frac{\rho_t}{\rho_\infty}}\|^2\rho_\infty dx,
$$
see below.
Suppose a LSI as in (\ref{LSI}) holds for $\mu_\infty(dx)=\rho_\infty(x)dx$. Let $f^2=\rho_t/\rho_\infty$ which indeed satisfies 
$$\int f^2d\mu=\int \frac{\rho_t}{\rho_\infty}\rho_\infty dx=1.$$
We then get
\begin{eqnarray}\nonumber
\mD(\rho_t\|\rho_\infty)=\int \frac{\rho_t}{\rho_\infty}\log\left(\frac{\rho_t}{\rho_\infty}\right)\rho_\infty dx=\int f^2\log f^2 d\mu\le\frac{1}{2\lambda}\int\|\nabla f\|^2d\mu=\\\frac{1}{2\lambda}\int \|\nabla \sqrt{\frac{\rho_t}{\rho_\infty}}\|^2\rho_\infty dx=\frac{1}{2\lambda}\int \|\nabla \log\left(\sqrt{\frac{\rho_t}{\rho_\infty}}\right)\|^2\rho_t dx=\frac{1}{8\lambda}\int \|\nabla \log\left(\frac{\rho_t}{\rho_\infty}\right)\|^2\rho_t dx \label{LSI'}
\end{eqnarray}
From (\ref{Decay}) and (\ref{LSI'}), we finally get
\begin{equation}\label{expdecay}
\frac{d}{dt}\mD(\rho_t\|\rho_\infty)\le -4\lambda\mD(\rho_t\|\rho_\infty)
\end{equation}
which implies exponentially fast decay of the relative entropy to zero. Thus $\rho_t$ converges in the (strong) entropic sense to $\rho_\infty$ and therefore in $L^1$. Thirty years ago Bakry and Emery proved that if the function $R$ is {\em strongly convex}, i.e.\ the Hessian of $R$ is uniformly bounded away from zero, then  $\rho_\infty$ satisfies a suitable LSI. This result has, since then, been extended in many ways, most noticeably by Villani \cite{Vil2}.

To establish entropic convergence of $\rho_t$ to $\rho_B$ for our degenerate diffusion model (\ref{OU}), we would need a suitable LSI of the form
$$\frac{1}{2}\int_{\mR^{n}}\int_{\mR^n}\langle
\Sigma\Sigma^\prime \nabla_v\log\frac{\rho_t}{\rho_B},\nabla_v\log\frac{\rho_t}{\rho_B}\rangle\rho_t\,dxdv\ge4\lambda \mD(\rho_t\|\rho_B).
$$ 
It is apparent that the possibility of establishing such a result depends only on the properties of the potential function $V$. Recently, some results in this directions have been reported in \cite{Bau} under various assumptions including the rather strong one that the Hessian of $V$ be bounded.

\section{Optimal steering to a steady state and reversibility}\label{sec:optimal}
Consider again the system of stochastic oscillators (\ref{OU}) and let $\bar{\rho}$, given by
\begin{equation}\label{steadystate}\bar{\rho}(x,v)=\bar Z^{-1}\exp\left[-\frac{H(x,v)}{kT_{{\rm eff}}}\right],
\end{equation}
be a desired thermodynamical state with $T_{{\rm eff}}<T$, $T$ being the temperature of the thermostat. Consider the controlled evolution
\begin{subequations}\label{CL12}
\begin{eqnarray}\label{CL1}
dx(t)&=&v(t)\,dt,\quad x(t_0)=x_0 \;{\rm a.s.}\\\label{CL2}
Mdv(t)&=&-Bv(t)\,dt-Uv(t)dt-\nabla V(x(t))dt+\Sigma dW(t), \quad v(t_0)=v_0 \; {\rm a.s.},
\end{eqnarray}
\end{subequations}
where $B$ and $\Sigma$ satisfy (\ref{FD}) and $U$  is a constant $n\times n$ matrix. We have the following fluctuation-dissipation relation
which is a direct consequence of Proposition \ref{Pro1}. 

\begin{cor}\label{prop:prop2} Under condition {\em (\ref{FD})}, the probability density $\bar{\rho}(x,v)$ in {\em (\ref{steadystate})} is invariant for the controlled dynamics {\em (\ref{CL12})} if and only if the following relation holds
\begin{equation}\label{FD2}
\frac{T-T_{{\rm eff}}}{T}\Sigma\Sigma^\prime=kT_{{\rm eff}}\left[U+U'\right].
\end{equation}
\end{cor}
Observe that $U$ satisfying (\ref{FD2}) always exist. For instance, if we require $U$ to be symmetric, it becomes unique and it is explicitly given by
\begin{equation}\label{optsymm}
U_{{\rm sym}}=\frac{1}{2}\left[\frac{T-T_{{\rm eff}}}{kTT_{{\rm eff}}}\Sigma\Sigma'\right].
\end{equation} 
Considerations on uniqueness, ergodicity and convergence are completely analogous to those of the Subsection \ref{generalcase} and will not be repeated here. We shall just assume that the potential function $V$ is such that an LSI for $\bar{\rho}(x,v)dxdv$ can be established \cite{Bau} leading to entropic exponential convergence of  $\rho_t(x,v)$ to $\bar{\rho}(x,v)$ for any $U$ satisfying (\ref{FD2}). Thus, such a control $-M^{-1}Uv$ achieves asymptotically the desired cooling.

{

It is interesting to investigate which of the feedback laws $-Uv$ which satisfy (\ref{FD2}) and therefore drive the system (\ref{CL12}) to the desired steady state $\bar{\rho}$, does it more efficiently. Following \cite[Section II-B]{CGP2}, we consider therefore the problem of minimizing the expected input power (energy rate)
\begin{equation}\label{power}
J_p(u)=\E\left\{u'u\right\}
\end{equation}
over the set of admissible controls
\begin{equation}\label{admissible}
{\mathcal U}_p=\left\{u(t)=-M^{-1}Uv(t) \mid U \mbox{ satisfies } (\ref{FD2})\right\}.
\end{equation}
Observe that, under the distribution $\bar{\rho}dxdv$, $x$ and $v$ are independent. Moreover, $\E\{vv'\}=kT_{{\rm eff}}M^{-1}$. Hence
$$\E\left\{u'u\right\}=\E\left\{v'U'M^{-2}Uv\right\}=kT_{{\rm eff}}\tr\left[M^{-1}U'M^{-2}U\right]. 
$$

We now proceed with a variational analysis that allows identifying the form of the optimal control. Let $\Pi$ be a symmetric matrix and consider the Lagrangian function
\begin{equation}
 \mathcal{L}(U,\Pi)=kT_{{\rm eff}}\tr\left[M^{-1}U'M^{-2}U\right]+\tr\left(\Pi(kT_{{\rm eff}}\left[U+U'\right]-\frac{T-T_{{\rm eff}}}{T}\Sigma\Sigma^\prime)\right)
 \end{equation}
 which is a simple quadratic form in the unknown $U$. Taking variations of $U$, we get
$$
\delta \mathcal{L}(U,\Pi;\delta U)=kT_{{\rm eff}}\tr\left(\left(M^{-1}\delta U' M^{-2}U+M^{-1}U'M^{-2}\delta U+\Pi\delta U+\Pi\delta U'\right)\right).
$$
Setting $\delta \mathcal{L}(U,\Pi;\delta U)=0$ for all variations, which is a sufficient condition for optimality, we get $M^{-2}UM^{-1}=\Pi$ which implies that $M^{-1}U$ equals the symmetric matrix $M\Pi M$. Thus, for an extremal point $U^*$, we get the {\em symmetry condition}
\begin{equation}\label{optgain}
U^*M^{-1}=M^{-1}(U^*)'.
\end{equation}
This optimality condition can be related to reversibility in the steady state. Indeed, repeating the analysis of Subsection \ref{reversibility} with $B+U$ in place of $B$, we get that the phase-space process (\ref{OU}) is reversible with the steady state distribution (\ref{steadystate}) if and only if
$$\Sigma\Sigma'=2kT_{{\rm eff}}(B+U).
$$
If we have reversibility in equilibrium, namely $B$ is symmetric positive definite satisfying (\ref{FD'}), we get
\begin{equation}\label{reversiblestst}U=\frac{T-T_{{\rm eff}}}{T_{{\rm eff}}}B=U_{{\rm sym}}>0.
\end{equation}
We collect these observations in the following result.
\begin{cor} Assume $M=mI_n$ a scalar matrix. Then $U^*=U_{{\rm sym}}$. This, under the assumption that $B$ satisfies {\em (\ref{FD'})}, is equivalent to reversibility in the steady state {\em (\ref{steadystate})}. If, morever, $B=m\beta I_n$ and $\Sigma=m\sigma I_n$, writing $T_{{\rm eff}}=(\beta/\beta+\gamma) T$, we get $U^*=m\gamma I$.
\end{cor}
It follows, in particular, that what is implemented in various applications \cite{LMC,BBP,Vin} does indeed minimize the expected control power (\ref{power}) among those satisfying (\ref{FD2}).

\newcommand{\cred}{\color{red}}
\section{Fast cooling for the system of stochastic oscillators}\label{sec:cooling}

Consider now the same system of stochastic oscillators (\ref{OU}) subject to an
external force represented by the control action $u(t)$:
\begin{subequations}\label{OU34}
\begin{eqnarray}\label{OU3}
dx(t)&=&\phantom{-}v(t)\,dt,\\\label{OU4}
Mdv(t)&=&-Bv(t)\,dt+u(t)dt-\nabla V(x(t))dt+\Sigma dW(t),
\end{eqnarray}
\end{subequations}
with $x(t_0)=x_0$ and $v(t_0)=v_0$ a.s.
Here $u$ is  to be specified  by the controller in order to achieve  { the desired cooling at a {\em finite} time $t_1$}. That is, we seek to steer the system of stochastic oscillators to the desired steady state $\bar{\rho}$ given in \eqref{steadystate} in finite time. Let $\mathcal U$ be the family of {\em adapted},\footnote{That is, the control process is ``causally dependent'' on the $(x,v)$ process.} {\em finite-energy} control functions such that the initial value problem (\ref{OU34}) is well posed on bounded time intervals and such that the probability density of the ``state'' process
\[
\xi^u(t_1)=\left(\begin{array}{c}x\\v\end{array}\right)
\]
is given by \eqref{steadystate}.
More precisely, $u\in\mathcal U$ is such that  $u(t)$ only depends on $t$ and on $\{\xi^u(s); t_0\le s\le t\}$ for each $t>t_0$, satisfies
$$\E\left\{\int_{t_0}^{t_1}u(t)' u(t) \,dt\right\}<\infty,
$$
and is such that $\xi^u(t_1)$ is distributed according to $\bar{\rho}$.
The family $\mathcal U$ represents  here the {\em admissible} control inputs which achieve the desired probability density transfer from $\rho_0$ to $\rho_1=\bar{\rho}$. Thence, we formulate the following {\em Schr\"{o}dinger Bridge Problem:}
\begin{problem}\label{formal1} Determine whether ${\mathcal U}$ is non-empty and if so, find
$
u^*:= {\rm argmin}_{u\in \mathcal U} \,J(u)$
where
\begin{equation}\label{index}
J(u):=\E\left\{\int_{t_0}^{t_1} \frac{1}{2}u(t)'\left(\Sigma\Sigma'\right)^{-1}u(t) \,dt\right\}.
\end{equation}
\end{problem}

The original motivation to study these problems comes from {\em large deviations of the empirical distribution} \cite{F2,ellis,DZ}, namely a rather abstract probability question first posed and, to some extent, solved by Erwin Schr\"{o}dinger in two remarkable papers in 1931 and 1932 \cite{S1,S2}. The solution of the large deviations problem, in turn, requires solving a {\em maximum entropy problem} on path space where the uncontrolled evolution plays the role of a ``prior" \cite{F2,W}, see also \cite{PT1,PT2,GP}. The latter, as we show in this specific case in Appendix \ref{sec:relative}, leads to Problem \ref{formal1}. Observe that, after $u^*$ has steered the system to $\bar{\rho}$ at time $t_1$, we simply need to switch to a control $u(t)=-Uv(t)$, with $U$ satisfying (\ref{FD2}), to keep the system in the desired steady state, see Section \ref{sec:example} for an illustrating example.

To simplify the writing here and in Appendix \ref{sec:relative}, we take $M=mI_n$, $B=m\beta I_n$ and $\Sigma=m\sigma I_n$ in (\ref{OU}): 
\begin{subequations}\label{OUS}
\begin{eqnarray}\label{OUS1}
dx(t)&=&v(t)\,dt,\quad x(t_0)=x_0 \;{\rm a.s.}\\\label{OUS2}
dv(t)&=&-\beta v(t)\,dt-\frac{1}{m}\nabla_x V(x(t))dt+\sigma dW(t), \quad v(t_0)=v_0 \; {\rm a.s.}
\end{eqnarray}
\end{subequations}
As we are now working on a finite time interval, the assumption that $B$ be a diagonal, positive definite matrix is not as crucial as it was in the previous two sections. Next we outline the variational analysis in the spirit of Nagasawa-Wakolbinger \cite{W} to obtain a result of Jamison \cite{Jam} for our degenerate diffusion (\ref{OUS}). Let $\varphi(x,v,t)$ be any positive, space-time harmonic function for the uncontrolled evolution, namely $\varphi$ satisfies on $\mR^{2n}\times [t_0,t_1]$
\begin{equation}\label{sth}\frac{\partial\varphi}{\partial t}+v\cdot\nabla_x\varphi+(-\beta v-\frac{1}{m}\nabla_x V)\cdot\nabla_v\varphi+\frac{\sigma^2}{2}\Delta_{v}\varphi=0.
\end{equation}
It follows that $\log\varphi$ satisfies
\begin{equation}\label{logeq}
\frac{\partial\log\varphi}{\partial t}+v\cdot\nabla_x\log\varphi+(-\beta v-\frac{1}{m}\nabla_x V)\cdot\nabla_v\log\varphi+\frac{\sigma^2}{2}\Delta_{v}\log\varphi=-\frac{\sigma^2}{2}\|\nabla_v\log\varphi\|^2.
\end{equation}
Observe now that, in view of (\ref{RE1}) in Appendix \ref{sec:relative}, the maximum entropy problem is equivalent to minimizing over admissible measures $P_u$ {on the space of paths} the functional
\begin{equation}\label{funct}I(P_u)=\E_{P_u}\left[\int_{t_0}^{t_1}\frac{1}{2\sigma^2}u\cdot u dt-\log\varphi(x(t_1),v(t_1),t_1)+\log\varphi(x(t_0),v(t_0),t_0)\right]
\end{equation}
since the endpoints marginals at $t=t_0$ and $t=t_1$ are fixed. Under $P_u$, by Ito's rule \cite{KS},
\begin{eqnarray}\nonumber d\log\varphi(x(t),v(t),t)&=&\frac{\partial\log\varphi}{\partial t}+v\cdot\nabla_x\log\varphi+(-\beta v-\frac{1}{m}\nabla_x V +u)\cdot\nabla_v\log\varphi\\
&&+\frac{\sigma^2}{2}\Delta_{v}\log\varphi(x(t),v(t),t)dt+\nabla_v\log\varphi(x(t),v(t),t)\sigma dW(t).\nonumber
\end{eqnarray}
Using this and (\ref{logeq}) in (\ref{funct}), we now get
\begin{align}\nonumber
I(P_u)&= \E_{P_u}\left[\int_{t_0}^{t_1}\frac{1}{2\sigma^2}u\cdot u dt-\log\varphi(x(t_1),v(t_1),t_1)+\log\varphi(x(t_0),v(t_0),t_0)\right]\\\nonumber
&=\E_{P_u}\left[\int_{t_0}^{t_1}\left(\frac{1}{2\sigma^2}u\cdot u\right.\right.\\\nonumber
&\hspace*{-20pt}\left.\left.-\left[\frac{\partial\log\varphi}{\partial t}+v\cdot\nabla_x\log\varphi+(-\beta v-\frac{1}{m}\nabla_x V +u)\cdot\nabla_v\log\varphi+\frac{\sigma^2}{2}\Delta_{v}\log\varphi\right](x(t),v(t),t) \right)dt\right.\\\nonumber
&\hspace*{15pt}\left.-\int_{t_0}^{t_1}\nabla_v\log\varphi(x(t),v(t),t)\sigma dW(t)\right]\\\nonumber
&=\E_{P_u}\left[\int_{t_0}^{t_1}\left(\frac{1}{2\sigma^2}u\cdot u-u\cdot\nabla_v\log\varphi(x(t),v(t),t)+\frac{\sigma^2}{2}\|\nabla_v\log\varphi(X_t,t)\|^2 \right)dt\right]\\
&=\E_{P_u}\left[\int_{t_0}^{t_1}\frac{1}{2\sigma^2}
\|u-\sigma^2\nabla_v\log\varphi(x(t),v(t),t)\|^2dt\right],\label{eq:IPu}
\end{align}
where we have used the fact that the stochastic integral has zero expectation. Then the form of the optimal control follows
\begin{equation}\label{optcontr}
u^*(t)=\sigma^2\nabla_v\log\varphi(x(t),v(t),t).
\end{equation}
Thus $u^*$ is in {\em feedback} form, so that the optimal solution is a Markov process as we know from the general theory \cite{Jam}. If for some $\varphi$ the closed-loop system \eqref{OU34} with control \eqref{optcontr} and initial distribution $\rho_0$ does satisfy the terminal distribution $\bar\rho$, that is, the solution $\rho(x,v,t)$ of the Fokker-Planck equation 
    \begin{equation}\label{eq:fokkerplanck}
    \frac{\partial \rho}{\partial t}+v\cdot\nabla_x\rho+\nabla_v\cdot \left[\left(-\beta v-\frac{1}{m}\nabla_x V+u\right)\rho\right]-\frac{\sigma^2}{2}\Delta_{v}\rho=0
    \end{equation}
with initial value $\rho(x,v,t_0)=\rho_0(x,v)$ satisfies the final condition $\rho(x,v,t_1)=\bar\rho(x,v)$,
then this control $u^*$ solves Problem \ref{formal1}. 
Let 
\[
    \hat{\varphi}(x,v,t)=\frac{\rho(x,v,t)}{\varphi(x,v,t)}.
\]
Then a long but straightforward calculation shows that $\hat\varphi$ satisfies a Fokker-Planck equation and we  obtain the system
\begin{subequations}\label{eq:schrodingersys}
\begin{eqnarray}\frac{\partial\varphi}{\partial t}+v\cdot\nabla_x\varphi+(-\beta v-\frac{1}{m}\nabla_x V)\cdot\nabla_v\varphi+\frac{\sigma^2}{2}\Delta_{v}\varphi=0,\\\frac{\partial\hat{\varphi}}{\partial t}+v\cdot\nabla_x\hat{\varphi}+\nabla_v\cdot \left[\left(-\beta v-\frac{1}{m}\nabla_x V\right)\hat{\varphi}\right]-\frac{\sigma^2}{2}\Delta_{v}\hat\varphi=0,
\end{eqnarray}
with boundary conditions
\begin{equation}
\varphi(x,v,t_0)\hat{\varphi}(x,v,t_0)=\rho_0(x,v),\quad \varphi(x,v,t_1)\hat{\varphi}(x,v,t_1)=\bar\rho(x,v).
\end{equation}
\end{subequations}
The system of linear equations with nonlinear boundary couplings (\ref{eq:schrodingersys}) is called the {\em Schr\"{o}dinger system}. Conversely, if a pair $(\varphi,\hat{\varphi})$ satisfies the Schr\"{o}dinger system \eqref{eq:schrodingersys}, then $P_{u^*}$ is the solution of the Schr\"{o}dinger bridge problem. Existence and uniqueness\footnote{The solution is actually unique up to multiplication of $\varphi$ by a positive constant $c$ and division of $\hat{\varphi}$ by the same constant.} for this system was guessed by Schr\"{o}dinger himself and proven in various degrees of generality by Fortet, Beurlin, Jamison and F\"{o}llmer \cite{Beu60, For, Jam74, F2}, see also \cite{CGP6.15} for a recent different approach. Hence, there is a unique control strategy $u^*$ in Problem \ref{formal1} that minimizes the control effort \eqref{index}. The optimal evolution steering the stochastic oscillator from $\rho_0$ to $\bar\rho(x)$ with minimum effort is given by
\begin{eqnarray*}\label{opt1}
dx(t)&=&v(t)\,dt,\\\label{opt2}
dv(t)&=&-\beta v(t)\,dt-\frac{1}{m}\nabla_x V(x(t))dt+ \sigma^2\nabla_v\log\varphi(x(t),v(t),t)dt+\sigma dW(t),
\end{eqnarray*}
where $\varphi$ solves together with $\hat{\varphi}$ the Schr\"{o}dinger system (\ref{eq:schrodingersys}). We observe that $-\sigma^2\log\varphi(x,v,t)$ plays the role of an artificial potential generating the external force which achieves the optimal steering.

\section{The case of a quadratic potential
}\label{sec:quadratic}
We consider the same situation as in Subsection \ref{quadratic} where the potential $V$ is simply given by the quadratic form
$$V(x)=\frac{1}{2}x'Kx,
$$
with $K$ a symmetric, positive definite $n\times n$ matrix. The dynamics of the stochastic oscillator (\ref{OU}) become linear and we can directly apply the results of \cite{CGP}. This is precisely the situation considered in \cite{Vin,BCD}. We proceed to show that it is possible to design a feedback control action which takes the system to the desired (Gaussian) steady state
$$\bar{\rho}=\bar Z^{-1}\exp\left[-\frac{\frac{1}{2}mv'v+\frac{1}{2}x'Kx}{kT_{{\rm eff}}}\right]$$
at the {\em finite} time $t_1$. The uncontrolled dynamics (\ref{OUS}) 
\begin{eqnarray}\label{L1}
dx(t)&=&\phantom{-}v(t)\,dt,\\\label{L2}
dv(t)&=&-\beta v(t)\,dt-\frac{1}{m}Kx(t)dt+\sigma dW(t).\nonumber
\end{eqnarray}
are in the form
$d\xi=\A\xi dt+\B dW$,
where
$$\xi=\left(\begin{matrix}x\\v\end{matrix}\right),\quad \A=\left(\begin{matrix}0&I\\-\frac{1}{m}K&-\beta I\end{matrix}\right), \quad \B=\left(\begin{matrix}0\\\sigma I\end{matrix}\right).
$$
Notice that the pair $(\A,\B)$ is controllable. Once again, introducing a control input $u(t)$, we want to minimize
$$\E\left\{\int_{t_0}^{t_1} \frac{1}{2}u(t)'u(t) \,dt\right\}
$$
under the controlled dynamics
\begin{subequations}
\begin{eqnarray}\label{C1}
dx(t)&=&v(t)\,dt,\\\label{C2}
dv(t)&=&-\beta v(t)\,dt-\frac{1}{m}Kx(t)dt+u(t)dt+\sigma dW(t),\end{eqnarray}
\end{subequations}
with $x(t_0)=x_0$, and $v(t_0)=v_0$ a.s.
Then, applying  \cite[Proposition 2]{CGP}, we get that the optimal solution is
\[
u^*(t)=-\sigma\B'\Pi(t)\xi
\]
where $(\Pi(t),{\rm H}(t))$ is the solution to the following system of Riccati equations
\begin{equation}\label{R1}
\dot{\Pi}(t)=-\A'\Pi(t)-\Pi(t)\A+\Pi(t)\B\B'\Pi(t),
\end{equation}
\begin{equation}\label{R2}
\dot{\rm H}(t)=-\A'{\rm H}(t)-{\rm H}(t)\A-{\rm H}(t)\B\B'{\rm H}(t),
\end{equation}
coupled through their boundary values by
\begin{eqnarray}
\frac{1}{kT}\diag\{K,\,mI\}&=&{\rm H}(t_0)+\Pi(t_0)\\
\frac{1}{kT_{\rm eff}}\diag\{K,\,mI\}&=&{\rm H}(t_1)+\Pi(t_1).
\end{eqnarray}
Because control effort is required to steer the system to a lower-temperature state, $\Pi(t)$ will be non-vanishing throughout. The precise form of the optimal control is in \cite[Theorem 8]{CGP}.

\section{Example}\label{sec:example}

This is an academic example, based on the linear model
\begin{eqnarray*}
dx(t)&=&\phantom{-}v(t)dt\\
dv(t)&=& -v(t)dt -x(t)dt+u(t)dt +dw(t)
\end{eqnarray*}
which corresponds to taking $m=1$, $\beta=1$, $\sigma=1$, and $K=1$ in suitable units. The goal is to steer and maintain the system starting from an intial temperature (in consistent units) of $T=\frac12$ to a final temperature $T_{\rm eff}=\frac{1}{16}$.

Thus, we seek an optimal $u(t)$ as a time-varying linear function of $\xi=(x,v)'$ to steer the system
from a normal distribution in phase space with zero mean and covariance
\[
{kT}\diag\{K,\,mI\}^{-1}=\frac12\left[\begin{matrix} 1&0\\0&1\end{matrix}\right]
\]
to a final distribution with zero mean and covariance
\[
{kT_{\rm eff}}\diag\{K,\,mI\}^{-1}=\frac{1}{2^4}\left[\begin{matrix} 1&0\\0&1\end{matrix}\right],
\]
over the time window $[0,\,1]$. Thereafter, the distribution of $x(t)$ remains normal maintaining the covariance via a choice of $u(t)$ which is a linear, time-invariant function of $v(t)$, namely $u(t)=-Uv(t)$, with now the scalar constant $U$ {satisfying (\ref{FD2})}. The figures show the trajectories of the inertial particles in phase space as a function of time and the respective control effort. Thus, Figure \ref{fig:1} shows typical sample paths and Figure \ref{fig:2} shows the nature of the corresponding control inputs.
The transition is effected optimally, using time-varying control, whereas at $t_1=1$, the value of the control switches to {the time-invariant linear function of $v(t)$ which maintains} thereafter the distribution of $(x(t),v(t))$ at the desired level.
}

\begin{figure}\begin{center}
\includegraphics[width=0.47\textwidth]{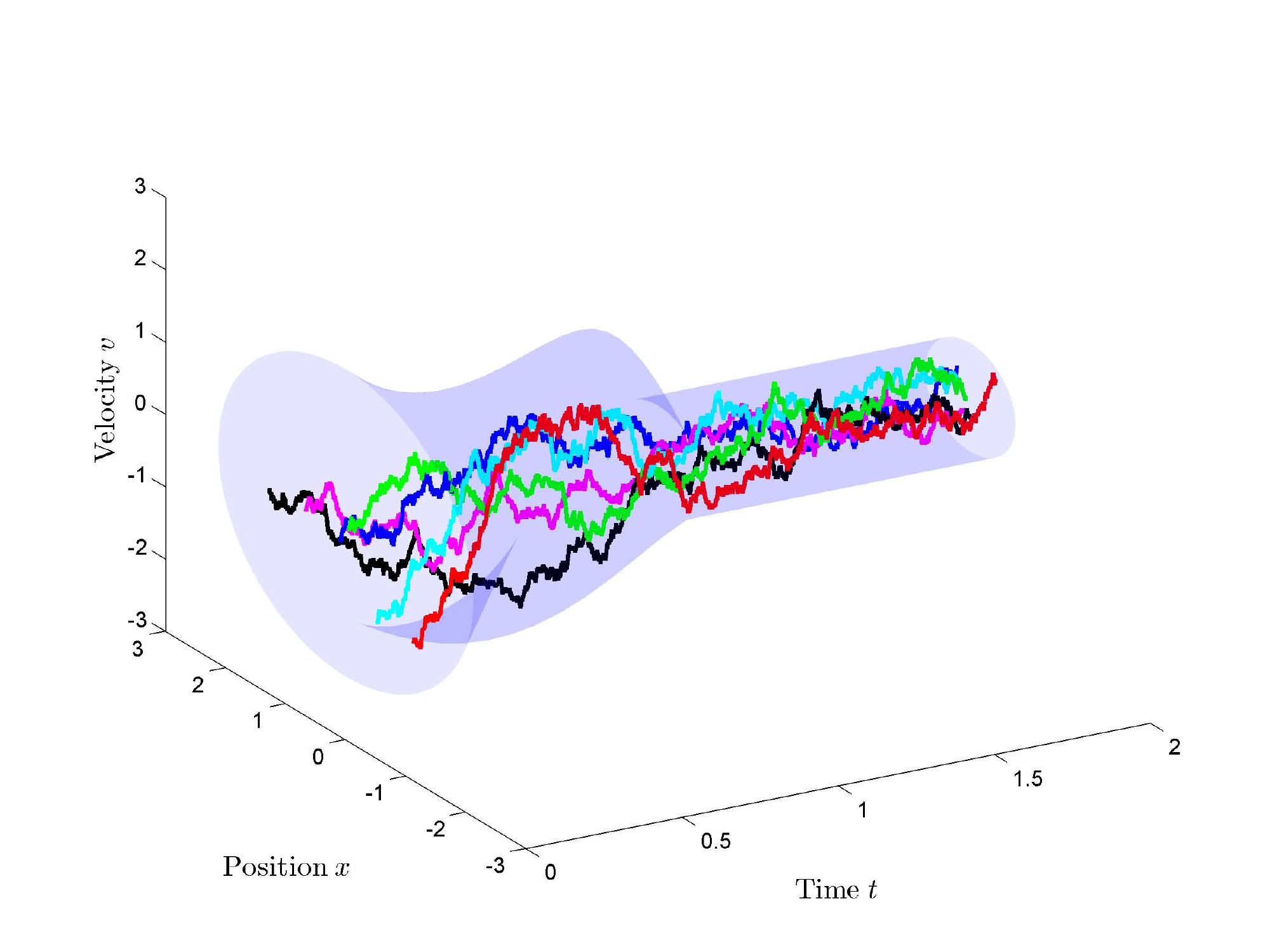}
   \caption{Inertial particles: trajectories in phase space}
   \label{fig:1}
\end{center}\end{figure}
\begin{figure}\begin{center}
\includegraphics[width=0.47\textwidth]{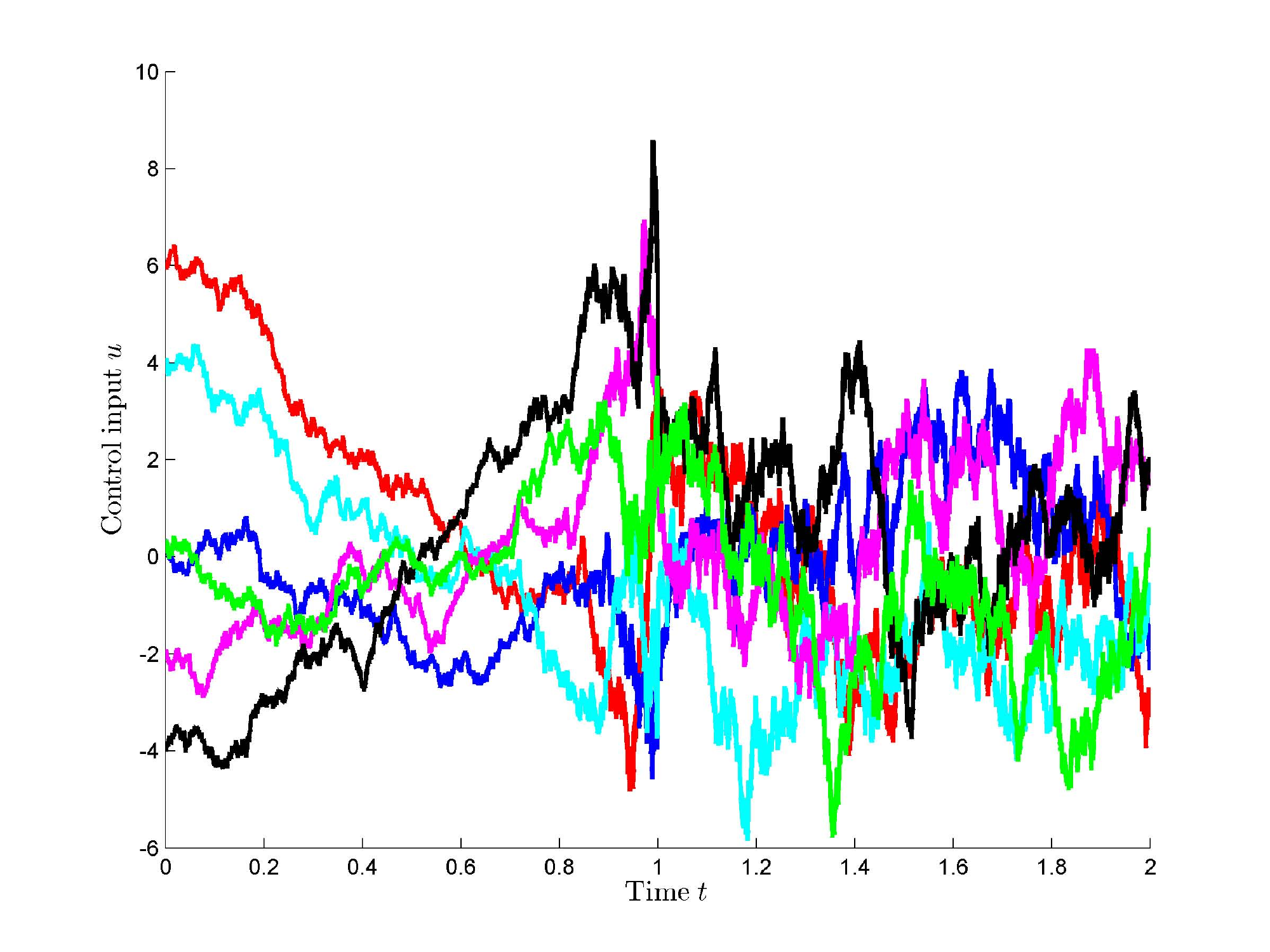}
   \caption{Inertial particles: control effort $u(t)$}
   \label{fig:2}
\end{center}\end{figure}

\appendix

\section{Relative entropy for stochastic oscillators measures}\label{sec:relative}

Consider
the same set up as in Section \ref{sec:cooling} and let $\D$ denote the space of probability measures on path spaces for phase space processes. Consider the process with Ito's differential
\begin{eqnarray}\nonumber
dx(t)&=&v(t)\,dt+ n^{-1/2}dZ(t),\quad x(t_0)=x_0 \;{\rm a.s.}\\\nonumber
dv(t)&=&-\beta v(t)\,dt-\frac{1}{m}\nabla V(x(t))dt+ u(t)+\sigma dW(t), \quad v(t_0)=v_0 \;{\rm a.s.},\nonumber
\end{eqnarray}
and let $P_u^n\in\D$ be the measure corresponding to a choice of a specific control law $u\in\mathcal U$.
Here $Z$ is standard $n$-dimensional Wiener processes independent of $W$ and of the initial conditions $x_0$, $v_0$. The difference with respect to the model in (\ref{OUS}) is that now we have also a ``weak" noise $n^{-1/2}dZ(t)$ affecting the configurational variables. Let
$$\Theta^2=\diag \left(\frac{1}{n},\sigma^2\right)
$$
denote the diffusion coefficient matrix for the above model. Next, using Girsanov's theorem \cite{IW,KS}, we compute the Radon-Nikodym derivative $\frac{dP_u^n}{dP_0^n}$ between the probability laws for the controlled and the uncontrolled (i.e., with $u=0$) processes.

Let $W_0$ be a Wiener measure starting with distribution $\rho_0(x,v)dxdv$ of $(x_0,v_0)$  at $t=t_0$. Since $W_0$, $P_u^n$ and $P_0^n$ have the same initial marginal, we get
\begin{eqnarray}\nonumber
\frac{dP_u^n}{dW_0}=\exp\left[\int_{t_0}^{t_1}\Theta^{-1}\beta_t^{P_u^n}\cdot \Theta^{-1}dX_t-\int_{t_0}^{t_1}\frac{1}{2}\beta_t^{P_u^n}\cdot\Theta^{-2}\beta_t^{P_u^n} dt\right],\quad P_u^n \; {\rm a.s.},\\ \frac{dW_0}{dP_0^n}=\exp\left[-\int_{t_0}^{t_1}\beta_t^{P_0^n}\cdot \Theta^{-1}dX_t+\int_{t_0}^{t_1}\frac{1}{2}\beta_t^{P_0^n}\Theta^{-2}\beta_t^{P_0^n}dt\right],\quad P_0^n \; {\rm a.s.}\Rightarrow P_u^n \; {\rm a.s.}.\nonumber
\end{eqnarray}
Therefore,
\begin{eqnarray}\nonumber\frac{dP_u^n}{dP_0^n}&=&\exp\left\{\int_{t_0}^{t_1}\left(\Theta^{-1}\beta_t^{P_u^n}-\Theta^{-1}\beta_t^{P_0^n}\right)\cdot\Theta^{-1}\left(\begin{matrix}dx_t\\dv_t\end{matrix}\right)+\frac{1}{2}\int_{t_0}^{t_1}\left[\beta_t^{P_0^n}\cdot\Theta^{-2}\beta_t^{P_0^n}-\beta_t^{P_u^n}\cdot\Theta^{-2}\beta_t^{P_u^n}\right]dt\right\}\\\nonumber&=&\exp\left\{\int_{t_0}^{t_1}\left(\Theta^{-1}\beta_t^{P_u^n}-\Theta^{-1}\beta_t^{P_0^n}\right)\cdot\left(\begin{matrix}dZ_t\\dW_t\end{matrix}\right)+\frac{1}{2}\int_{t_0}^{t_1}\left(\beta_t^{P_u^n}-\beta_t^{P_0^n}\right)\cdot\Theta^{-2}\left(\beta_t^{P_u^n}-\beta_t^{P_0^n}\right) dt\right\}\\\nonumber&=&\exp\left\{\int_{t_0}^{t_1}\left[\Theta^{-1}\left(\begin{matrix}0\\u\end{matrix}\right)\right]\cdot\left(\begin{matrix}dZ_t\\dW_t\end{matrix}\right)+\frac{1}{2}\int_{t_0}^{t_1}\left(\begin{matrix}0\\u\end{matrix}\right)\cdot\Theta^{-2}\left(\begin{matrix}0\\u\end{matrix}\right)dt\right\}\nonumber\\\nonumber&=&\exp\left\{\int_{t_0}^{t_1}\frac{1}{\sigma}u\cdot dW_t+\int_{t_0}^{t_1}\frac{1}{2\sigma^2} u\cdot u dt.\right\}.
\end{eqnarray}
We observe that this Radon-Nikodym derivative does not depend on $n$. 

Now, let $P_u$ and $P_0$ be the measures in $\D$ corresponding to the situation when there is no noise in the position equation (i.e., $n=\infty$). In this case, as expected,
$$\frac{dP_u}{dP_0}=\int_{t_0}^{t_1}\frac{1}{\sigma}u\cdot dW_t+\int_{t_0}^{t_1}\frac{1}{2\sigma^2}u\cdot u dt.
$$
To derive this formula, one could have also resorted to a general form of Girsanov's theorem \cite[Thm 4.1]{IW}, \cite[(5.3)]{Fischer}.
Assuming that the control satisfies the finite energy condition
$$\E\left[\int_{t_0}^{t_1} u\cdot u dt\right]<\infty,
$$
the stochastic integral
$$\int_{t_0}^{t_1}\frac{1}{\sigma}u\cdot dW_t
$$
has zero expectation. We then obtain that the relative entropy between $P_u$ and $P_0$ is 
\begin{equation}\label{RE1}
\mD (P_u\|P_0)=\E_{P_u}\left[\log\frac{dP_u}{dP_0}\right]=\E_{P_u}\left[\int_{t_0}^{t_1}\frac{1}{2\sigma^2}u\cdot u dt\right],
\end{equation}
which is precisely the index in Problem \ref{formal1} in the case $\Sigma=\sigma I_n$.

\end{document}